\documentclass[aps,preprint,preprintnumbers,amsmath,amssymb,nofootinbib]{revtex4}
\usepackage{amssymb}
\usepackage{lscape}
\usepackage{amsmath}
\usepackage{bm}
\usepackage{graphicx}
\usepackage{epsfig}
\begin{document}
\title{Mass density of the Earth from a Gravito-Electro-Magnetic 5D vacuum}
\author{$^{2}$ Juan Ignacio Musmarra, $^{1,2}$ Mariano Anabitarte,  $^{1,2}$ Mauricio Bellini
\footnote{{\bf Corresponding author}: mbellini@mdp.edu.ar} }
\address{$^1$ Departamento de F\'isica, Facultad de Ciencias Exactas y
Naturales, Universidad Nacional de Mar del Plata, Funes 3350, C.P.
7600, Mar del Plata, Argentina.\\
$^2$ Instituto de Investigaciones F\'{\i}sicas de Mar del Plata (IFIMAR), \\
Consejo Nacional de Investigaciones Cient\'ificas y T\'ecnicas
(CONICET), Mar del Plata, Argentina.}

\begin{abstract}
We calculate the mass density of the Earth using a Gravito-Electro-Magnetic theory on an extended 5D Schwarzschild-de Sitter metric, in which we define the vacuum.
Our results are in very good agreement with that of the Dziewonski-Anderson model.
\end{abstract}
\maketitle

\section{Introduction}

A detail description of the interior of the Earth, like that
of the other planets, remains the biggest topic in the
earth sciences and cosmochemistry. The knowledge of the composition
of the core provides clues into the nature of the
Earth’s electric and magnetic fields, gravitational characteristics,
heat flux values, and has important geodynamic
implications\cite{Litasov}. The Earth can be divided by composition into the crust, mantle, and core.
The crust consists mainly of igneous and metamorphic rocks with a relatively thin layer of sediment and sedimentary rocks on top.
Crust comes in two varieties: continental and oceanic. Continental crust consists mainly of felsic rocks such as granite, and is about $30-50$ {\em km} thick, varying from place to place. The oceanic crust consists mainly of somewhat more mafic rocks such as basalt, and is about $5-10$ {\em km} thick.
The mantle consists of a further $2890$ kilometers of denser ultramafic rocks. The core is the innermost part of the Earth, having a radius of $3400$ {\em km}. It can be divided into the outer core, which is molten, and the inner core, with a radius of $1220$ {\em km}, which is solid. The core is made mainly of iron. As with the existence of the crust, this can be explained on the hypothesis of differentiation: just as the light substances rose to the top, so the denser substances would sink to the bottom.

The study of the Earth density is an important issue that has been studied, for instance, using neutrino collisions with the Earth matter\cite{samp}. There are experimental date in this sense provided by the IceCube experiments\cite{ice}. In this work we shall explore the inner mass distribution of the Earth, using the Gravito-Electro-Magnetic theory from a 5D vacuum. This theory was introduced, but in a cosmological context in\cite{mb}, and later extended using the Weitzenb\"ock geometry\cite{rb}. This is a gravito-electrodynamic formalism
constructed with a penta-vector with components $A_b$ that can be applied to any physical system in the framework of the induced matter theory (IMT)\cite{stm}, which is supported in the
Campbell-Magaard theorem\cite{cm}. This theorem serves as a ladder to move between manifolds whose dimensionality
differs by one. Furthermore, is valid in any number of dimensions ($D\geq 2$), so that every solution of the 4D Einstein equations with arbitrary energy momentum tensor can
be embedded, at least locally, in a solution of the 5D Einstein field equations in vacuum.

\section{4D Schwarzschild-de Sitter induced from a 5D vacuum.}

In order to describe the mass and charge densities: ($\rho_{m}$, $\rho_{q}$), we shall suppose that the Earth is isotropic. Therefore both will be only dependent of $r$.
To describe a charged sphere with mass, we shall use a 5D extended Reissner-Nordstr\"on-de Sitter metric, but with mass and charge distributions that are outside the inner horizon
\begin{equation}
^{(5)}ds^{2}=(\frac{\psi}{\psi_{0}})^{2}[f(r) \, dt^{2} - \frac{1}{f(r)} \, dr^{2} - r^{2} \, d \Omega^{2}]-d \psi^{2}.
\end{equation}
Here, $\psi$ is the non-compact extra dimension and $d\Omega^2=(d\theta^2+\sin^2\theta\, d\phi^2$ and
\begin{equation}
f(r)=1-\frac{2 \, G \, \rho_{M}(r) \, r^{2}}{c^{2}}+\frac{G \, \rho_{q}(r)^{2} \, r^{4}}{c^{4}}-\frac{1}{3} \, \Lambda \, r^{2}.
\end{equation}
Since this kind of metric is canonical, it is guaranteed the 4D connections are given by the 5D connections evaluated on
the foliation $\psi=\psi_{0}$. Furthermore, we shall require that the Einstein equations describe a 5D vacuum: $G_{ab}=0$. This condition
is fulfilled for the following differential equation
\begin{equation}
r\,\frac{d \rho_{M}(r)}{dr} +3 \rho_{M}(r) = r^3 \rho_{q}(r) \frac{d \rho_{q}(r)}{dr} + \frac{5}{2} r^2 \,\rho^2_{q}(r) + \frac{3}{2 \psi^2_0},
\end{equation}
which has the general solution
\begin{equation}\label{rho}
\rho_{q}(r)= \pm \frac{1}{3} \frac{\sqrt{3}\sqrt{Gr(\Lambda r^{3}c^{4} \psi_{0}^{2} + 6 \rho_{M}(r) c^{2} r^{3} G \psi_{0}^{2} - 3 r^{3}c^{4} + 3 C_{1} G \psi_{0}^{2})}}{Gr^{3} \psi_{0}}.
\end{equation}
Here, $C_{1}$ is a constant of integration. If we replace (\ref{rho}) in $f(r)$, we obtain
\begin{equation}\label{fr}
f(r)=1+\frac{\bar{M}\,G}{c^{4}r}-\frac{r^{2}}{\psi_{0}^{2}}.
\end{equation}
This expression  corresponds to an effective 4D Schwarzschild metric for the choice: $\psi_{0}=\sqrt{\frac{3}{\Lambda}}$, $\bar{M}=0$ and a Schwarzschild-de Sitter one for
$\psi_{0}=\sqrt{\frac{3}{\Lambda}}$, $\bar{M}=-2Mc^{2}$. The last choice describes a 5D punctual mass $M$ of an object with a cosmological constant $\Lambda$.

\section{Gravitational Potential}

\begin{figure}[htpb]
	\centering
    \includegraphics[scale=1]{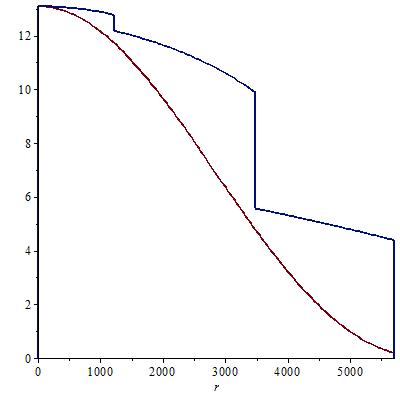}
	\caption{Radius [$km$] vs. density [$\frac{kg}{m^{3}}10^{-3}$]. The red curve describe the Dziewonski-Anderson model and the blue curve shows our theoretical predictions without consider the layers of the inner structure }
  \label{fig:ima2}
\end{figure}

\begin{figure}[htpb]
	\centering
    \includegraphics[scale=1]{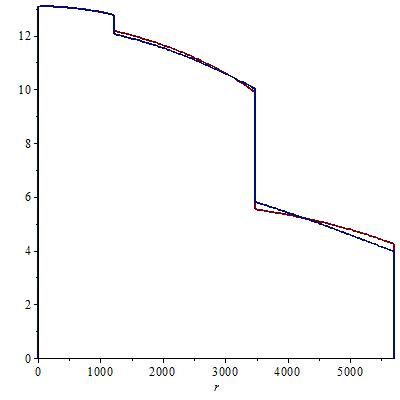}
	\caption{Radius [$km$] vs. density [$\frac{kg}{m^{3}}10^{-3}$]. The red curve describe the Dziewonski-Anderson model and the blue curve shows our theoretical predictions for $C_{5}=C_{6}=6.54425$ in the inner core, $C_{5}=C_{6}=6.2395$ in the outer core and $C_{5}=C_{6}=3.423$ at the mantle. In all the cases was considered $s=s_0=0.00012$.}
  \label{fig:ima4}
\end{figure}

We shall consider a 5D physical system described by both, gravitational and electric fields, in a vacuum. The expression obtained in (\ref{fr}) assures that the system
is in a 5D vacuum: $G_{ab}=0$. The density Lagrangian for relativistic electric and gravitational fields which are stationary, is
\begin{equation}\label{act}
{\cal L}=\frac{1}{f(r)}(\vec{E}\cdot\vec{E}-\vec{\nabla}\Phi\cdot\vec{\nabla}\Phi),
\end{equation}
where $\vec{E}=-\vec{\nabla}\Theta$ is the 5D electric field and $\Phi$ is the 5D gravitational potential. For free fields, the resulting dynamics is
\begin{equation}
\begin{array}{l}
\displaystyle ^{(5)}\Box\Phi=0,\\
\displaystyle ^{(5)}\Box\Theta=0,\\
\end{array}
\end{equation}
such that $^{(5)}\Box$ represents the 5D D\'Alambertian operator and $\Theta$ the electric potential. We can propose the Fourier expansion
\begin{eqnarray}
\Phi(x^a) &=& \frac{1}{(2\pi)^{3/2}} \int ds \, \int d^3 k \left[ A_{lm\lambda s}(k) \,\Phi_{lm\lambda s}(x^{a}) + A^{\dagger}_{lm\lambda s}(k) \, \Phi^*_{klm\lambda s}(x^a)\right], \\
\Theta(x^a) &=& \frac{1}{(2\pi)^{3/2}} \int ds \, \int d^3 k \left[ B_{lm\lambda s}(k) \,\Theta_{lm\lambda s}(x^{a}) + B^{\dagger}_{lm\lambda s}(k) \, \Theta^*_{klm\lambda s}(x^a)\right].
\end{eqnarray}
In order to obtain the solution we shall propose
\begin{eqnarray}
\Phi_{lm\lambda s}(x^a)& = & R^{(g)}_{ls\lambda}(r) \, Y_{lm}(\theta, \varphi) \, T^{(g)}_{\lambda}(t) \, \Psi^{(g)}_{s}(\psi), \\
\Theta_{lm\lambda s}(x^a)& = & R^{(e)}_{ls\lambda}(r) \, Y_{lm}(\theta, \varphi) \, T^{(e)}_{\lambda}(t) \, \Psi^{(e)}_{s}(\psi).
\end{eqnarray}
where the superscripts $^{(g)}$, $^{(e)}$, denote the gravitational and electric nature of the solutions and
\begin{eqnarray}
\frac{d^{2} T^{(g,e)}_{\lambda}(t)}{dt^{2}}+ \lambda^{2} T^{(g,e)}_{\lambda}(t)=0, \label{e1} \\
\frac{\partial Y_{lm}(\theta, \varphi)}{\partial \theta}\frac{\cos(\theta)}{\sin(\theta)}+\frac{\partial^{2} Y_{lm}(\theta, \varphi)}{\partial \theta^{2}}+\frac{1}{\sin^{2}(\theta)}Y_{lm}(\theta, \varphi)-l \, (l+1) \, Y_{lm}(\theta, \varphi)=0, \label{e2} \\
\frac{4}{\psi}\frac{d \Psi^{(g,e)}_{s}(\psi)}{d \psi}+ \frac{d^{2} \Psi^{(g,e)}_{s}(\psi)}{d^{2} \psi}+ s^{2}\Psi^{(g,e)}_{s}(\psi)=0, \label{e3} \\
\frac{\psi_{0}^{2}\lambda^{2} R^{(g,e)}_{ls\lambda}(r)}{\psi^{2}f(r)} - \frac{4r}{\psi^{2}}\frac{dR^{(g,e)}_{ls\lambda}(r)}{dr}
-\frac{2\psi_{0}^{2}G\mu}{\psi^{2}r^{2}c^{2}}\frac{dR^{(g,e)}_{ls\lambda}(r)}{dr}+\frac{2\psi_{0}^{2}}{\psi^{2}r}\frac{dR^{(g,e)}_{ls\lambda}(r)}{dr} \nonumber \\
+\frac{(\psi_{0}^{2}-r^{2})}{\psi^{2}}\frac{d^{2} R^{(g,e)}_{ls\lambda}(r)}{d^{2} r}-\frac{2\psi_{0}^{2}G\mu}{\psi^{2}c^{2}r}\frac{dR^{(g,e)}_{ls\lambda}(r)}{dr}-\frac{\psi_{0}^{2} l \, (l+1)}{\psi^{2}r^{2}}R^{(g,e)}_{ls\lambda}(r)+s^{2}R^{(g,e)}_{ls\lambda}(r)=0. \label{e4}
\end{eqnarray}
Here, $\lambda$, $l$ and $s$ are constant related the separation of variables. The equations (\ref{e1}), (\ref{e2}) and (\ref{e3}), support the solutions
\begin{eqnarray}
T^{(g,e)}_{\lambda}(t)=C_{1}\exp^{i\lambda t}+C_{2}\exp^{-i\lambda t} + C, \\
Y_{lm}(\theta, \varphi)=\sum_{l,m}\sqrt{\frac{(2l+1)(l-m)!}{4\pi(l+m)!}}\exp^{-im\varphi}P_{lm}(\cos(\theta)), \\
\Psi^{(g,e)}_{s}(\psi)=C_{3}\frac{(\cos(s\psi)s\psi-\sin(s\psi))}{\psi^{3}}+C_{4}\frac{(\cos(s\psi)+\sin(s\psi)s\psi)}{\psi^{3}}.
\end{eqnarray}
Because, we are searching spherically symmetric field solutions independent of the time, we shall impose $C_{1}=C_{2}=l=0$ and $C=1$. The relevant equations here is the (\ref{e4}),
which describes the radial behavior of the fields. In order to resolve the equation (\ref{e4}) will be sufficient to consider the approximation to a Schwarzschild solution due to the
fact we are interested in solutions outside the Schwarzschild radius and $r_{s} < r \ll \left( \frac{M\,G \psi_0}{c^4}\right)^{1/3}$. The general solution for the approximated differential
equation for $R^{(g,e)}_{ls\lambda}$, is
\begin{eqnarray}
R^{(g,e)}(r)|_{\Lambda \approx 0}&=&C_{5}\exp^{\sqrt{-\lambda^{2}\psi_{0}^{2}-\psi^{2}s^{2}}\frac{r}{\psi_{0}}}{\cal HC}\left[\alpha,\beta,\gamma,\delta,\eta,z(r)\right](-c^{2}r+2GM)^{i\frac{2M G\lambda}{c^{2}}} \nonumber \\
&+& C_{6}\exp^{\sqrt{-\lambda^{2}\psi_{0}^{2}-\psi^{2}s^{2}}\frac{r}{\psi_{0}}}{\cal HC}\left[\alpha,\beta,\gamma,\delta,\eta,z(r)\right](-c^{2}r+2GM)^{-i\frac{2MG\lambda}{c^{2}}},
\label{rad}
\end{eqnarray}
where ${\cal HC}\left[\alpha,\beta,\gamma,\delta,\eta,z\right]$\footnote{The confluent Heun fucntion ${\cal HC}\left[\alpha,\beta,\gamma,\delta,\eta,z(r)\right]$ is a solution of the differential equation
\begin{displaymath}
\frac{d^2{\cal HC}}{dz^2} + \frac{\left[\alpha  z^2 + (\beta +\gamma -\alpha+2) z - (\beta+1)\right]}{z(z-1)}
\frac{ {\cal HC}}{dz} + \frac{1}{2} \frac{\left[ [ (\beta+\gamma+2)\alpha + 2 \delta ] z -(\beta+1) \alpha +
(\gamma+1)\beta + 2\eta +\gamma\right] }{z(z-1)} {\cal HC}=0.
\end{displaymath}}
is the confluent Heun function with parameters $\alpha$, $\beta$, $\gamma$, $\delta$, $\eta$ and argument $z$\cite{heun}. In our case
these parameters and the argument are given by
\begin{eqnarray}
\alpha&=&-\frac{4GM\sqrt{-\lambda^{2}\psi_{0}^{2}-\psi^{2}s^{2}}}{\psi_{0}c^{2}},\nonumber \\
\beta &= & i\frac{4GM\lambda}{c^{2}}, \nonumber \\
\gamma& = & 0, \nonumber \\
\delta & = &-\frac{4G^{2}M^{2}(2\lambda^{2}\psi_{0}^{2}+\psi^{2}s^{2})}{c^{4}\psi_{0}^{2}}, \nonumber \\
\eta & = & \frac{((-l^{2}-l)c^{4}+8\lambda^{2}G^{2}M^{2})\psi_{0}^{2}+4G^{2}M^{2}\psi^{2}s^{2}}{c^{4}\psi_{0}^{2}}, \nonumber \\
z(r) &= & 1-\frac{c^{2}r}{2GM}.
\end{eqnarray}
Hence, the modes $\Phi_{lm\lambda}(x^a)$ and $\Theta_{lm\lambda}(x^a)$ will be given by
\begin{eqnarray}
\Phi_{lm\lambda} & = &\int_{-\infty}^{\infty}ds\int_{-\infty}^{\infty}d\lambda \, A(s,\lambda) \, R^{(g)}_{ls\lambda}(r) \, Y_{lm}(\theta, \varphi) \, \Psi^{(g)}_{s}(\psi), \\
\Theta_{lm\lambda} & = & \int_{-\infty}^{\infty}ds\int_{-\infty}^{\infty}d\lambda \, A(s,\lambda) \, R_{ls\lambda}(r) \, Y_{lm}(\theta, \varphi) \, \Psi^{(e)}_{s}(\psi).
\end{eqnarray}
When we make a static foliation such that we take $\psi=\psi_0$, the resulting value for the parameter $s$ will be $s=s_0$, as determined by eq. (\ref{e3}). Therefore, the
induced gravitational potential $\overline{\Phi}$\footnote{We denote with a ''bar'' the induced gravitational potential in order to no cause confusion with the
5D field $\Phi_{klm}$.}
\begin{equation}
\overline{\Phi}_{lm\lambda}(t,r,\theta,\varphi)=\int_{-\infty}^{\infty}d\lambda \, \int ds \, A(s,\lambda) \, R^{(g)}_{ls\lambda}(r) \, Y_{lm}(\theta, \varphi) \, T^{(g)}_{\lambda}(t) \, \Psi^{(g)}_{s}(\psi_{0})\,
\delta\left(s-s_0\right).
\end{equation}
It is expected that when we make the foliation on the extra dimension $\psi$, one should recover the dynamics of the 4D space-time
\begin{equation}
^{(4)}\Box\overline{\Phi}(x^{\alpha})=4\pi G \, \overline{\rho}_M(x^{\alpha}), \label{4d}
\end{equation}
where $\overline{\rho}_M(x^{\alpha})$ is the mass density on the induced 4D hypersurface. If we compare the 5D vacuum dynamical equation on the 5D vacuum:
$^{(5)}\Box\Phi=0$, with (\ref{4d}), is reasonable to propose
\begin{equation}
\Phi(r,\theta,\varphi,t,\psi_{0}) \sim \overline{\rho}_M(r,\theta,\varphi,t).
\end{equation}
In order to estimate the volumetric mass density, we shall neglect the time dependence: $\lambda=0$. Furtermore, we shall suppose that the density has spherical symmetry: $l=0$. With respect to the confluent Heun functions, we shall require that takes the value: $\left.{\cal HC}\left[\alpha,\beta,\gamma,\delta,\eta,z(r)\right]\right|_{r=r_s}=1$, at the Schwarzschild radius. This radius is very small for the Earth: $r_s=8.9 \times 10^{-3}\,\, {\rm mts.}$, so that we shall take the mass density as maximum at $r=r_s$. In order to the solution comply with this conditions we adjust the coefficients $C_{5}=C_{6}=6.54425$ in (\ref{rad}). Finally, we shall require that the value of the mass density be canceled at $r=6371 \times 10^3\,{\rm mts}$. With this condition we adjust the parameter $s=s_0 = 0.00025$. We can make the comparison with the Dziewonski-Anderson model\cite{Anderson}, which is based on seismic date of the Earth between $r=0$ and $r=5701 \times 10^3\,{\rm mts}$. As one can see in the
fig. (\ref{fig:ima2}), the results are not good. This is because we are not taken into account the different geologic layers of the Earth in its interior. If this is done, due to the fact the equation (\ref{4d}) is linear, we can apply the superposition principle. In order to do it, we must modify the boundary conditions. As in the previous case we shall assume that $C_{5}=C_{6}$.
In the fig. (\ref{fig:ima4}) we have plotted the mass density of the Earth jointly with the Dziewonski-Anderson curve, using $C_{5}=C_{6}=6.54425$ in the inner core, $C_{5}=C_{6}=6.2395$
in the outer core and $C_{5}=C_{6}=3.423$ at the mantle. In all the cases was considered $s=s_0=0.00012$. We can see that these calculations agree very good with that of the
Dziewonski-Anderson model, which were obtained using seismic date\cite{Anderson}. A very important fact is that these mass distributions are incompatible with a charge monopole distribution in (\ref{rho}). For this reason the inner Earth would be neutral from the point of view of the electric charge.

\section{Final Comments}

We have studied a model to calculate the mass density of the Earth using a Gravito-Electro-Magnetic theory on a background given by an extended 5D Schwarzschild-de Sitter metric, in which we define the vacuum. In our approach, we have neglected the electro and magnetic dynamical sources, which should be important in a Gravito-Electro-Magnetic theory in order to calculate the magnetic field of the Earth. Notice that in the Lagrangian density considered in (\ref{act}), only we are taking into account stationary gravito-electric fields on a 5D vacuum. A very important fact is that there is only one $s$ mode related to the extra dimensional field that contributes to the general solution: $s=s_0=0.00012$. This value is the same in all the layers of the Earth, but with different values of $C_5=C_6$-values. However, the calculation for the distribution of the mass in the inner of the Earth by us obtained is in very good agreement with other models and date obtained in different contexts.


\section*{Acknowledgements}

\noindent The authors acknowledge CONICET (Argentina) and UNMdP for financial support.

\end{document}